\documentclass[16pt,aps,nofootinbib]{revtex4}
\usepackage{amsmath}
\usepackage{graphicx}

\newcommand{\Gammamat}{\boldsymbol \Gamma}
\newcommand{\Rmat}{\mathbf R}
\renewcommand{\vec}[1]{\boldsymbol #1}

\DeclareMathOperator{\Trace}{Tr}

\begin{document}

\title{Renormalisation  Group Flow and Kaon Condensation}
\author{Boris Krippa$^{1}$}
\affiliation{$^1$School  of Science and Technology,
Nottingham Trent University, NG1 4BU, UK}
\date{\today}
\begin{abstract}
Functional renormalisation group approach is applied to a system of kaons at finite chemical potential.
 A set of approximate flow equations for the effective couplings is derived and solved. At high scale the system is found to be at the normal phase whereas at some critical value of the running scale it undergoes the phase transition (PT) to the phase with a spontaneously broken symmetry with the kaon condensate as an order parameter. The value of the condensate turns out to be quite sensitive to the kaon-kaon scattering length.

\end{abstract}
\maketitle

The study of phase structure of dense hadron matter is one of the key problems of the strong interaction physics described by QCD.  At low energy  QCD can be reduced to an effective theory containing Goldstone bosons as the effective degrees of freedom. The extension of this effective theory to the  finite density requires taking into consideration nonzero isospin/strange chemical potential. The physical motivation for such an extension 
is provided by a wide range of phenomenon occurring in  heavy-ion collisions and the Equation of State  (EoS) of the neutron stars \cite{Lat}. At finite density these bosons may condense.  This condensation shows up in a variety of the physical systems. For example, there is a  possibility that a kaon condensation may exist in a core of the neutron stars so that a realistic analysis of such possibility as well as estimates of the value of the condensate may turn out to be important for establishing correct EoS.

It is rather well known that, depending on the value of the chemical potential, the following phases can be realised:  a normal phase with the trivial vacuum and  pion/kaon condensation. In this letter we focus on the phenomena of   kaon condensation.  More precisely, we will consider the case of kaon condensation as the competing phase for the Color Flavour Locked (CFL) phase of quark matter at high density. As was shown in \cite{Bed} when nonzero quark masses are introduced they act in a way analogous  to the applied chemical potential. The main conclusion of Ref. \cite {Bed} was that nonzero quark masses favor meson condensation so that more symmetric CFL phase is unlikely to occur in nature.  The corresponding term is proportional to the combination $M^{+}M/(2\mu)$ (where $M$ is the quark mass matrix and $\mu$ is the quark number chemical potential). This term makes the CFL unstable and lead to meson condensation. Due to a large strange quark mass  kaon condensation is more preferable then pion condensation. In more realistic scenario these two condensations may mix and result in a very complicated phase diagram. In this paper however we consider only somewhat idealised case of ``clean'' kaon condensation.  The more general case of the mixed pion/kaon condensation will be reported elsewhere. In contrast to kaon condensation in baryon matter  which involves the $K^-$ mesons condensation in our case the $K^-$  condensed phase is disfavored (for details see refs. \cite{Bed}, \cite{Kap}).

Our main analytic tool is  the functional renormalisation group (FRG),
the nonperturbative  approach making use of  the Legendre transformed  effective action: $\Gamma[\phi_c]=W[J]-J\cdot \phi_c$, where $W$ is the usual partition function in the presence of some external source $J$. The effective  action $\Gamma$ generates the one particle irreducible  Green functions and it reduces to the effective potential for the homogeneous systems. In the FRG  with the  effective action $\Gamma$ an artificial renormalisation group flow is introduced, generated by a momentum scale $k$ and  one can define   the
scale dependent effective action by integrating over components of the 
fields with $q \leq k$. The RG trajectory then interpolates between the 
classical effective action of the underlying field theory (defined at large $k$) when the quantum fluctuation effects are excluded, and the full quantum effective action (defined at $k=0$) with all quantum fluctuations taken into account. In the practical calculations the flow is generated by the cut-off function $R(k)$ which suppresses modes with $q \leq k$. This cut-off function goes to zero in the limit $k \rightarrow 0$ when the physical observables are extracted and behaves like $k^2$ at large scale. In the idealised case of the exact solution of the  flow equation the results are independent of the functional form or/and the parameters of the cut-off function. However the exact solution is not possible in all physically interesting systems  so that  only the approximate independence is possible in practical applications of the approach. 

The FRG method has been successfully applied to   different physical systems ranging from the dynamics of the cold atoms to  a quantum gravity and asymptotic safety. In the context of  phase transitions the approach  has been used to study the phase structure of neutron matter
\cite{Kri1}, the phenomena of a Bose-Einstein condensation (BEC)\cite{Wet1}, and both balanced \cite{Kri2} and imbalanced \cite{Kri3} strongly interacting many-fermion systems  etc.

The corresponding evolution equation for $\Gamma$ \cite{Wet2} in the FRG has a deceptively simple
one-loop structure  and can be written as
\begin{equation}
\partial_k\Gamma=-\frac{i}{2}\,\Trace \left[
(\Gammamat ^{(2)}_{BB}-\Rmat_B)^{-1}\,\partial_k\Rmat_B\right].
\label{eq:Gamevol}
\end{equation}
Here $\Gammamat ^{(2)}_{(BB)}$ is the matrix containing second
functional derivatives of the effective action with respect to the
boson fields and $\Rmat_{B}$ is a matrix containing the
corresponding cut-off functions which must vanish when the running scale approaches zero. The flow equation as written above is just a differential form of a general functional integral and has therefore a little practical sense. As has already been mentioned in order to actually solve the  flow equations and be able to analyse the physical system one needs to rely on some approximations/truncations. The common strategy is that one usually  chooses a certain ansatze for the effective action with a finite number of the couplings motivated by the form of the action at the starting scale and/or the underlying symmetries. Then the ansatze is substituted into the general flow equations which  can then be written in the form of a set of the partial differential equations for the couplings.

 The phase transition to a condensed phase occurs within the range of applicability of the effective theory describing the low-energy limit of QCD at nonzero strange potential so that the  ansatze for the effective action used in this work is motivated by such a theory, namely  the linear sigma model with a finite chemical potential

\begin{equation}
\Gamma[\phi,\phi^\dagger] = \int d^4x\,
\left[Z_\phi\,(\partial_0 + i \mu)\phi^\dagger\,(\partial_0 - i \mu)\phi
-Z_m\,\partial_i\phi^\dagger\partial_i\phi - U(\phi,\phi^\dagger)\right],
\end{equation}
where  $Z_{\phi}$ and $Z_m$ are the  renormalisation factors depending on the running scale and $\phi$ is a complex doublet field defined as follows
\begin{equation}
\phi = \frac{1}{\sqrt 2} \begin{pmatrix}
\phi_1 + i \phi_2 \\ \bar\phi_1 + i \bar\phi_2
\end{pmatrix}
\end{equation}
The first and second components of the doublet can be identified with the pair of ($K^+, K^0$) and ($K^-, \bar K^0$) mesons correspondingly. The effective potential depends only  on the combination $\rho = \phi^{\dagger}\phi$. We expand the effective potential $U(\rho)$ near its scale dependent minima  and keep terms up to the order $\rho^3$ having in mind a possibility of the first order phase transition.
\begin{equation}
U(\phi,\phi^\dagger)=  u_1(\rho-\rho_0)
+\frac{1}{2}\, u_2(\rho-\rho_0)^2 + \frac{1}{6}\, u_3(\rho-\rho_0)^3 + \bar u_1(\bar\rho-\bar\rho_0)
+\frac{1}{2}\, \bar u_2(\bar\rho-\bar\rho_0)^2 + \frac{1}{6}\, \bar u_3(\bar\rho-\bar\rho_0)^3 + . . . 
\label{eq:potexp}
\end{equation}
Here the first three terms correspond to the expansion near the minimum with respect to the first doublet and the rest is the expansion near the minimum with respect to the second doublet. Note that the standard mass term is included in the definition of the $u_1$ coupling. At large scale we observe symmetric state with the trivial minimum of the effective potential whereas at lower scale comparable with the value of chemical potential the symmetry becomes spontaneously broken and formation of the condensate is expected. 

The chemical potential is provided by the external conditions (we assume $\mu >$ 0). The action is invariant under the global $SU(2) \times U(1)$ group, where $SU(2)$ is the isospin group and $U(1)$ is related to hypercharge. It essentially captures the mean features of the kaon condensation phenomena. Substituting the assumed ansatze for $\Gamma$ into the general flow equation for the scale dependent effective action and performing the contour integration one can get the evolution equation for the effective potential which acts as a driving term generating the flow of the couplings. 

\begin{eqnarray}
\partial_k U
&=&\,\frac{1}{4}\int\frac{d^3{\vec q}}{(2\pi)^3}\,\frac{(2 Z_{\phi} Q^2_1 - \alpha - \beta - 2 R) \partial_k R}
{4 Z^2_{\phi} Q^3_1- 2 Z_\phi Q_1(\alpha + \beta +2 R + 4 \mu^2 Z_\phi)}\nonumber\\
\noalign{\vskip 5pt}
&&+\,\frac{1}{4}\int\frac{d^3{\vec q}}{(2\pi)^3}\,
\frac{(2 Z_{\phi} Q^2_2 - \alpha - \beta - 2 R) \partial_k R}
{4 Z^2_{\phi }Q^3_2- 2 Z_\phi Q_2(\alpha + \beta +2 R + 4 \mu^2 Z_\phi)}\label{eq:potevol}
\end{eqnarray}

where

\begin{equation}
\alpha = Z_m q^2 + u_1 +u_2 (3 \rho_1 +\rho_2 - \rho_0) + \frac{u_3}{2}( 4 \rho_1(\rho_1 + \rho_2 - \rho_0) + (\rho_1 + \rho_2- \rho_0)^2),
\end{equation}

and

\begin{equation}
\beta = Z_m q^2 + u_1 +u_2 ( \rho_1 + 3\rho_2 - \rho_0) + \frac{u_3}{2}( 4 \rho_2(\rho_1 + \rho_2 - \rho_0) + (\rho_1 + \rho_2- \rho_0)^2).
\end{equation}

Here $\rho_i = \phi_i^{\dagger}\phi_i$ and  the scale dependent factors $Q_1$ and $Q_2$  determine the pole positions of the corresponding propagator

\begin{equation}
Q_{1(2)} = \frac{1}{\sqrt{2}}\sqrt{\alpha + \beta +2 R + 4 \mu^2 Z_\phi) \pm \sqrt{\alpha + \beta +2 R + 4 \mu^2 Z_\phi)^2 - 4((\alpha +R)(\beta +R) - \alpha\beta)}},
\end{equation}
Another two poles are those with the overall minus sign in front. The positive and negative values of energy correspond to creation and annihilation of the exitations, respectively. For our purposes it is sufficient to pick up only the positive ones.

Note that the pole position defines the corresponding dispersion relations in the general case of nonzero regulator $R \ne$ 0 . Taking  $R\rightarrow 0$, $Z_\phi \rightarrow 1$
and $u_1\rightarrow$ 0 
  one can recover the dispersion relations in the broken phase  derived in Refs. \cite{Mir,Son}. 
\begin{equation}
Q^b_{1,2} = \sqrt{ 3 \mu^2 - m^2+ q^2) \pm \sqrt{ (3 \mu^2  - m^2)^2) + 4\mu^2 q^2 }},
\end{equation}

where $m$ is the kaon mass. Another dispersion relations, also derived in Refs. \cite{Son, Mir} in the limit of $R\rightarrow 0$, $Z_\phi \rightarrow 1$
 take the form 

\begin{equation}
\bar {Q}^b_{1,2} = \sqrt{\mu^2+ q^2} \pm  \mu.
\end{equation}

As one can easily be seen from the above expressions two of 
the dispersion relations describe Goldstone bosons, i.e gapless modes when $q \rightarrow$ 0. This is the example of the nontrivial situation when the number of the broken generators of the symmetry group is not equal to the number of the massless modes. Indeed, since the initial global $SU(2) \times U(1)$ symmetry is broken to $U(1)$ one could expect three massless modes. The problem was studied in details in \cite{Mir, Son, Leu, Nie} where is was shown that in the system with a broken Lorenz symmetry the number of gapless modes may be lesser then the number of broken generators. The physical reason is the presence of the chemical potential which induces  a mass splitting between the doublets ($K^+, K^0$) and ($K^-, \bar K^0$) so that  the first one acquire the effective mass $m - \mu$ whereas the effective mass  for the second one becomes $m + \mu$.

The  couplings may in general depend not only on running scale but also on the magnitude of the condensate so that we define the total derivative as 

\begin{equation}
d_k=\partial_k + (d_k\rho)\,\frac{\partial}{\partial\rho}  , 
\end{equation}
where $d_k\rho=d\rho/dk$. Applying this to the effective potential and neglecting higher order terms
 gives the set of the flow equations 
\begin{eqnarray}
-u_2\,d_k\rho_1 =\left.\frac{\partial}{\partial \rho_1}
\Bigl(\partial_k U\Bigr)\right|_{\rho_1=\rho_0},\\
\noalign{\vskip 5pt}
 d_k u_2 -u_3 d_k \rho_1 =\left.\frac{\partial^2}{\partial \rho^2}
\Bigl(\partial_k U\Bigr)\right|_{\rho_1=\rho_0}, \\
\noalign{\vskip 5pt}
 d_k Z_\phi = -\,\frac{1}{2}\left.\frac{\partial^3}{\partial^2 \mu\partial\rho}
\Bigl(\partial_k U\Bigr)\right|_{\rho_1=\rho_0},\\
\noalign{\vskip 5pt}
d_k u_3 =\left.\frac{\partial^3}{\partial\rho^3}
\Bigl(\partial_k U\Bigr)\right|_{\rho_1=\rho_0}
\end{eqnarray}

The actual flow is determined by the choice of the cut-off function $R$. We have chosen the cut-off in the form suggested in \cite{Lit} $ R(k,q) = (k^2 - q^2) \Theta (k-q).$ The advantage of this form is that it simplifies some algebra, allowing part of the calculations to be carried out analytically. One notes, however, that this advantage may disappear for more complicated cases like, for example, in the mixed boson-fermion systems. In this exploratory study we take into account the scale dependence of the renormalisation constant $Z_\phi$ which contributes to the evolution of the effective mass via the  $Z_{\phi}\mu^2\phi^\dagger\phi $ term and may therefore influence the position of the transition point from the symmetric to broken phase. On the contrary the   renormalisation factor $Z_m$ is not directly coupled to a mass term and does not really contribute to its evolution in symmetric phase. On top of that $Z_m$ itself evolves very slowly in symmetric case.  One notes that whereas running of $Z_m$ in the broken phase may somehow change the numerical value of the condensate  it is very unlikely that its evolution may change a general conclusion about the order of the phase transition. We therefore assume that $Z_m$ does not run at all and put $Z_m$ = 1 from now on.

To solve the system of the flow equations one needs to fix a set of initial conditions. They were determined as follows. One starts from the effective action in vacuum where we put $u^v_1(k=0) = m$, $Z_v(k=0)=1$,  $u^v_3(k=0) = 0$  and determine  $u^v_2(k=0)$ from  the  kaon-kaon scattering length in vacuum. The flow equations in vacuum can easily be obtained from the general expression for the $\partial_k U$ by differentiating it with respect to $\rho$ and $\mu$ and putting 
$\rho = 0$, $\mu=0$ afterwards. For example, the flow equations for the couplings $u^v_1$ and  $u^v_2$ 
take the form

\begin{equation}
\partial_k u^v_1(k) = \frac{k^4 u^v_{2}}{8\pi^2(k^2 + u^v_1(k))^{3/2}} 
\end{equation}

\begin{figure}
\begin{centering}
\includegraphics[width=14cm]{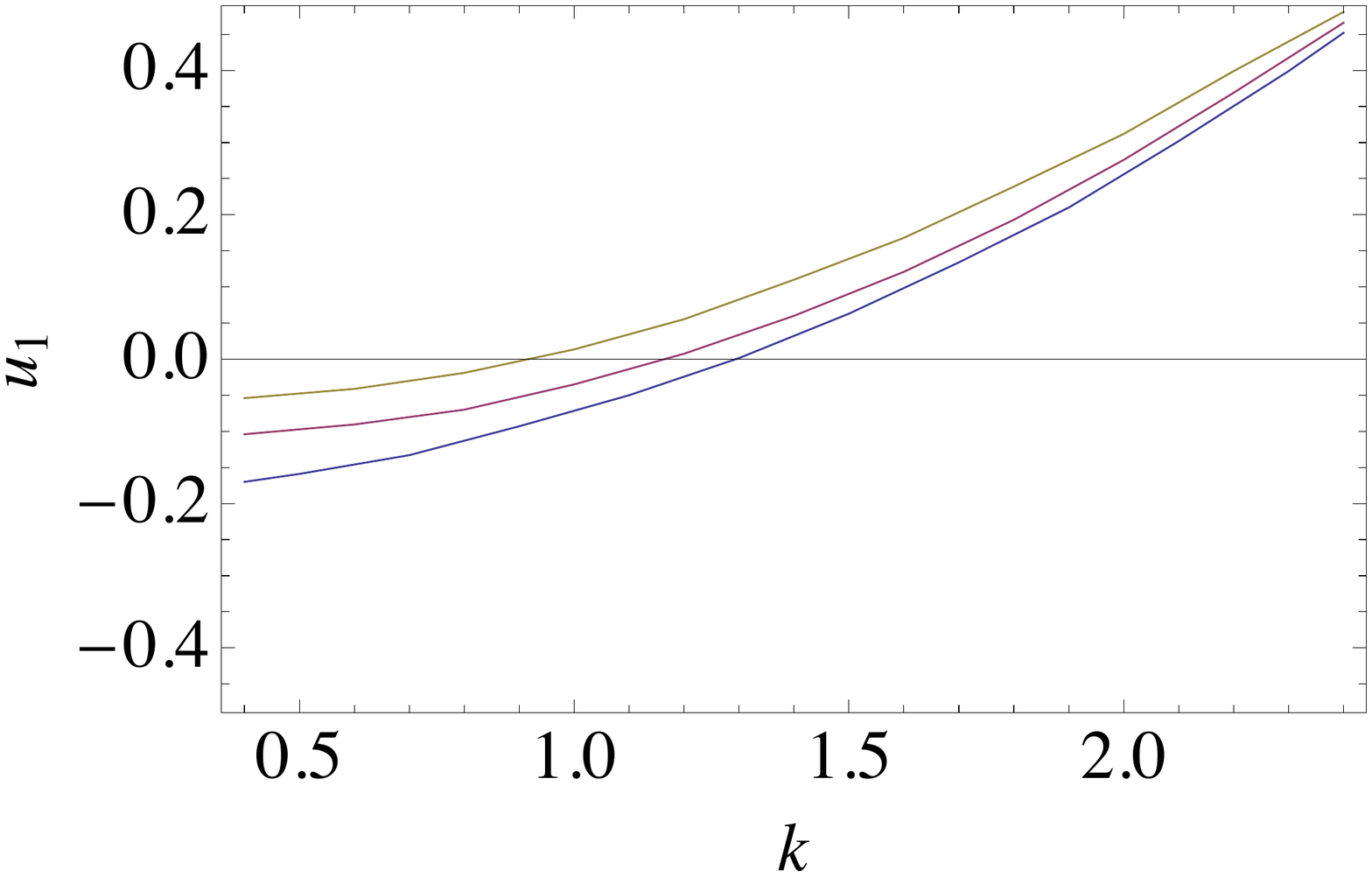}
\par\end{centering}
\vspace*{-55mm}
\caption{The evolution of the mass term with the scale. The upper (green online), medium  (orange online) and lower (blue online) curves correspond to $\mu$ = 0.51 GeV, 0.55 GeV and 0.595 GeV correspondingly.}

\end{figure}

and 

\begin{equation}
\partial_k u^v_2(k) =  \frac{k^4 (4 (k^2 + u^v_1) u^v_{3} - 5 (u^v_2)^2)}{16\pi^2(k^2 + u^v_1(k))^{5/2}}
\end{equation}

The flow equations for the other couplings can be obtained in a similar way. The value of $u^v_3(k)$ is found to be very small at $k = \Lambda$ compared to the other couplings and the renormalisation factor $Z_\phi$ does not really run in vacuum so that we put $Z_\phi = 1$ at starting scale.

These equations are solved using the values of couplings in vacuum at $k=0$ as initial conditions. From the vacuum flow   the couplings at the starting scale $k=\Lambda$  for the general  flow at finite density  are extracted. The value of the scale $\Lambda$ is chosen to be much larger then any other mass scale involved in the problem so that the finiteness of the chemical potential plays a little role there and therefore $u^v_1(\Lambda)\simeq u_1(\Lambda)$.
We have taken the value $\Lambda = 50$ GeV as the starting scale. It is large enough to provide practically independent results for the couplings at the physical scale $k = 0$. At the scale $k \sim \mu$ the mass term approaches zero thus signaling the onset of spontaneous symmetry breaking (SSB). The corresponding vacuum expectation value of the field $\phi$ becomes nonzero and the system undergoes phase transition.

We show on Fig.1 the behaviour of the mass term  in the vicinity of the critical scale at the several values of the chemical potential. The critical value of the chemical potential for such a transition is found to be at $\mu=m$ as it should. At the scale $k \simeq $ 3 GeV the curves merge and follow this pattern up to starting point.

The change of the transition point as a function of the chemical potential is such  that it grows with the increase of the chemical potential but, in all cases the system undergoes a transition to the broken phase at the scale $k \sim \mu$. At this scale the mass term vanishes   and the condensate develops. 
        
   On the other hand the character of the change  of the mass term for the other fields in $\bar\phi$ is such that it stays positive at any scale so that SSB never happens. One may, therefore conclude that, whereas $K^0$ and $K^+$ mesons condense,  the pair of the $\bar K^0$ and $K^-$ mesons do not condense in agreement with the results from \cite{Bed} and \cite{Kap}. One notes that the mechanism of the condensation considered in this paper and in Refs. \cite{Bed,Kap} is related to an instabilities  in the colour - flavour locked phase of QCD \cite{Alf} and is therefore quite different from the conventional condensation of $K^-$ mesons in nuclear matter suggested in \cite{Kap1} which is related to attractive interaction between $K^-$ mesons and nucleons.

\begin{figure}
\begin{centering}
\includegraphics[width=11.2cm]{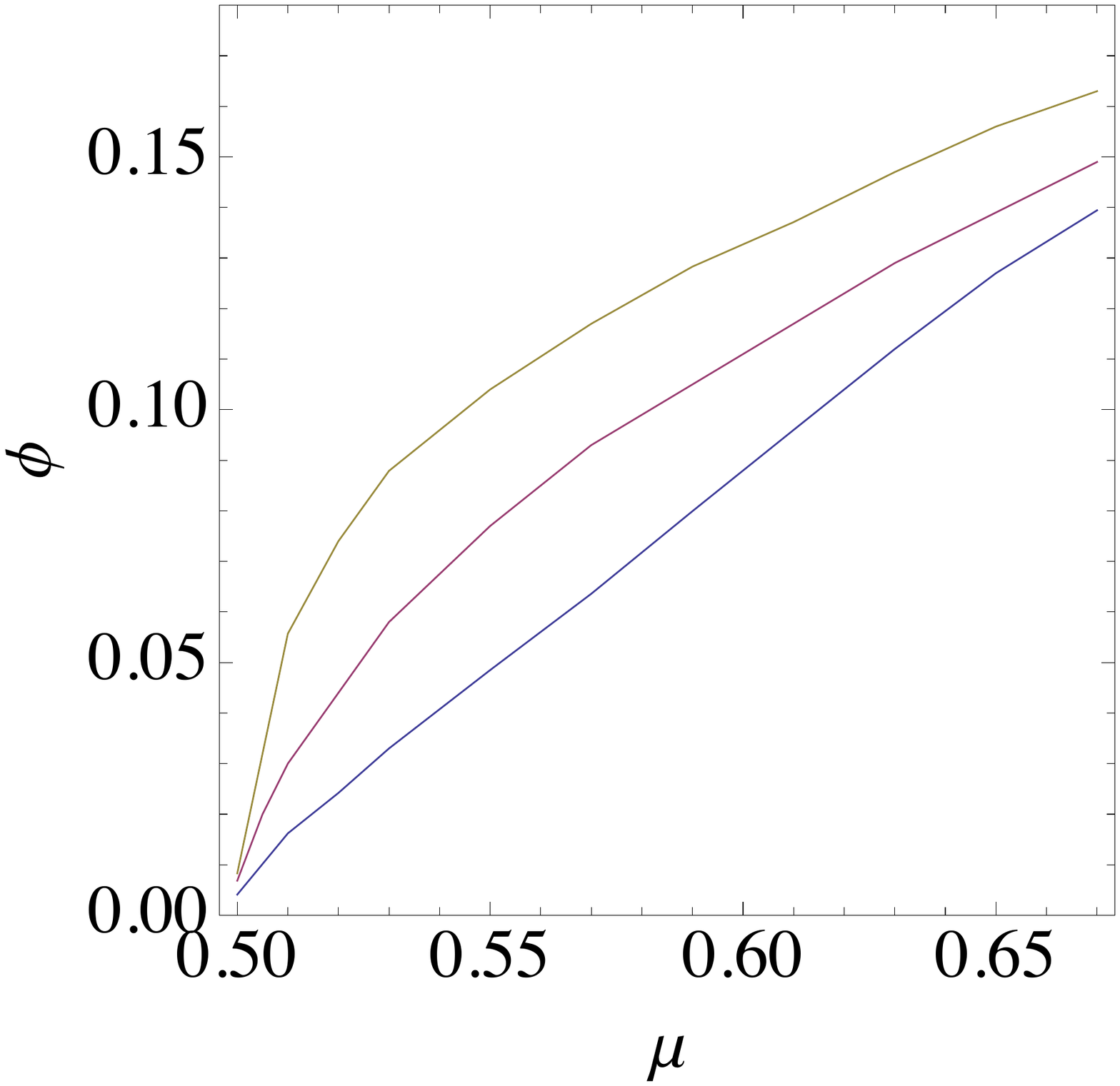}
\par\end{centering}
\vspace*{-12mm}
\caption{The  value of the condensate as the function of the chemical potential for the different values of the coupling $u_2$ corresponding to kaon-kaon scattering lengths from the lattice calculations \cite {Sas}(middle curve), lowest order chiral perturbation theory \cite{Bed} (lower curve) and phenomenological model \cite{Jap} (upper curve)} . 

\end{figure}
     
       In  Fig.2 we show the behaviour of the condensate as the function of chemical potential at physical scale when $k \rightarrow 0$ and for three values of the $u_2$ coupling. As one can see from Fig.2 the value of the condensate grows with the increase of the chemical potential in all three cases. This growth is relatively fast at the threshold $\mu \simeq \mu_{crit}$ and then slows down for the larger values of the chemical potential. We note however, that the results become less reliable with the departure of the chemical potential from its threshold value $\mu_c = m$ as  kaon  in this case may asquire the nonzero momentum and interact not only in the $S$ but also in $P$ - wave so that the corresponding coupling in the effective action must be related to the scattering amplitude at nonzero momentum rather then  just to the  scattering length. Therefore, the energy/momentum dependence of the couplings neglected in this study may become progressively more important at larger $\mu$.
   
  \begin{figure}
\begin{centering}
\includegraphics[width=15.6cm]{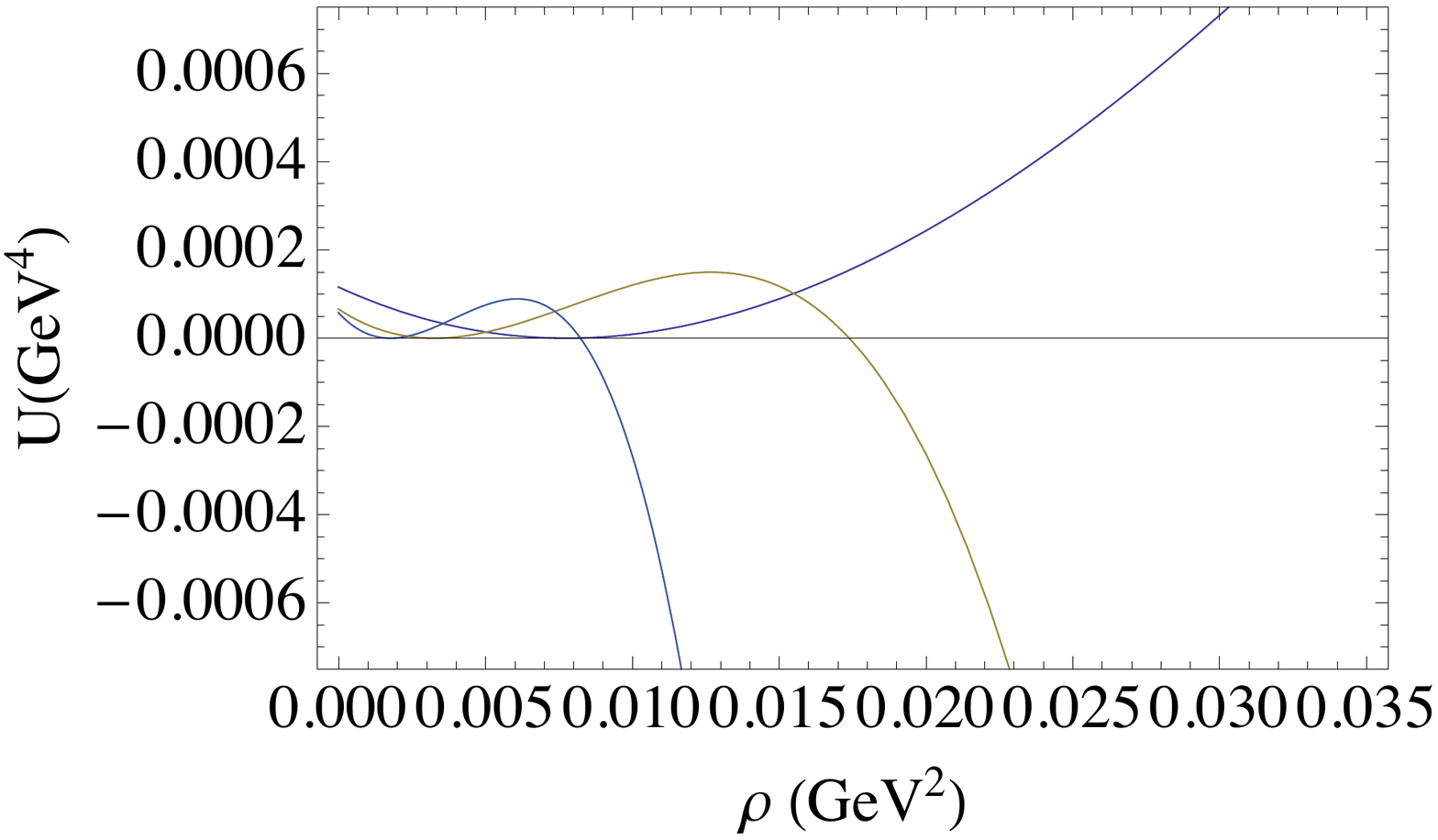}
\par\end{centering}
\vspace*{-5cm}
\caption{The  effective potential  for the different values of the coupling $u_2$ corresponding to kaon-kaon scattering lengths from the lattice calculations \cite {Sas}(pink  online), lowest order chiral perturbation theory \cite{Bed} (yellow  online) and phenomenological OBEP model \cite{Jap} (blue  online)} . 
\end{figure}

The important ingredient  of the approach is the running coupling $u_2$, which is related to the kaon-kaon scattering length. To get the results shown on the middle curve in Fig.2 we used the value for the scattering length obtained in the lattice calculations $a_{lat}$= -0.310$m^{-1}$ \cite{Sas}.  The lower curve corresponds to the use of the lowest order chiral perturbation theory ($\chi$PT) result $a = \frac{m^2}{16\pi f^2_{\pi}}$ \cite{Bed} and the upper one is obtained using the phenomenological model from Ref. \cite{Jap}. One can see from Fig.2 that the difference between the results  is quite noticeable, both in the threshold region and beyond and the largest value of the scattering length (obtained from $\chi$PT) leads to the smallest  value of the condensate.
It becomes less sensitive to the kaon-kaon interaction when   the chemical potential  becomes significantly larger then kaon mass  but  as argued above in this case using energy/momentum independent interaction vertices  may not be sufficient.
  Taking into account the next-to-leading order in $\chi$PT  \cite{Oll} does not make this difference any smaller. 
 On the contrary, it leads to even  larger suppression of the condensate  and a further deviation from the lattice results. In any case the estimates based on the $SU(3)$ version of chiral perturbation theory are not very reliable because of its rather slow convergence    and some sort of resummation techniques are probably required like those developed for the kaon-nucleon scattering \cite{Kri4}. 
                     
           One can in general conclude that the effects of the meson rescattering, often ignored in the mean field studies  turn out to be an important ingredient of the underlying dynamics, although the quantitative estimates rely on the knowledge of the kaon-kaon low-energy scattering amplitude which is not known well enough at the moment. It would be interesting to consider the same type of questions for the case of the pion condensation, where there are many available  experimental data on the low energy  pion-pion scattering amplitudes  and well developed theoretical approaches. The similar issue can also be explored in the case of meson condensation at both finite chemical potential and temperature although we expect a reduced sensitivity of the condensate value to meson rescattering effects in this case.
                     
            It is worth mentioning that a significant dependence of the condensate value on the boson-boson interaction is quite a general issue in the Bose-Einstein condensation theory and can be studied in  completely different physical settings. The cold atomic Bose gas in the BEC regime is one such example.
           
           In Fig.3 we show the behaviour of the effective potential as a function of $\rho$. One can see from the figure that  at larger values of the kaon-kaon scattering length the position of the minimum is shifted towards the origin. It is interesting  that this tendency gets in some sense ``saturated'' for some unnaturally large values of the scattering length so that the position of the minimum becomes independent of the further increase of the scattering length. It implies that the condensation always occurs for the chemical potentials larger then some critical one but the actual value of the condensate crucially depends on the values of the couplings involved. Whether or not it is true for the other types of meson condensations like pion condensation is an   open questions at the moment which would be very interesting to explore.  It should  in some sense be easier to perform a similar analysis in the case of the pion condensation as the pion-pion scattering length is known to much better accuracy.
           
           A somewhat naive explanation of this ``universality'' could be that when one of the dimensionful parameters (scattering length in our case) become much larger then the other scales involved one observes some pattern of approximate universality, similar to what has been found for many-fermion systems. Whether it is a manifestation of some underlying dynamics or just an artefact of the assumptions used to approximate the effective action remains an open question. This point clearly deserves a further study, based on more sophisticated ansatze for the effective action.
           
             The shape of the potential suggests that the phase transition is of second order regardless of the strength of the $u_2$ coupling (or kaon-kaon scattering length). We could not find any indication of the other possible local minimum, needed for the first order phase transition to take place. Certainly, to rule out this possibility completely one needs to solve the flow equation with an unexpanded effective potential. The corresponding calculations are much more involved and are currently in progress.
   
     In general one can conclude that the  $u_2$ coupling generates a significant source of the theoretical uncertainty related to a poor knowledge of the kaon-kaon scattering amplitude at threshold. On the other hand this quantity is important for establishing the correct Equation of State (E0S) for neutron stars.

In the present work we have used the FRG approach to analyse the phenomena of kaon condensation. We have found that at $\mu = m$ the system undergoes the second order phase transition to the broken phase with a nontrivial vacuum resulting in formation of the nonzero order parameter, the kaon condensate. However the exact value of the condensate can't at present be determined accurately because of lack of knowledge about the magnitude of the kaon-kaon scattering length. Whereas our ansatze used is in principle sufficient to see if the transition is  first order we have found no indication of that.

There are several ways of how the present approach can be extended and improved. First, the energy/momentum dependence of the interaction vertices can be taken into account. It may be important for a realistic estimation  of the depletion effect on the condensate. Second, instead of rather toy model for the effective action, used in this paper one can utilize a more realistic action based on the effective Lagrangian like that derived in Ref. \cite{Bed}. The induced or effective chemical potential considered there will emerge as the result of the RG flow. Third, physically interesting case of  competing chiral and kaon condensates can be also considered. Apart from that it would be interesting to explore a mixed system with both kaon and pion condensates. These condensates may influence each other via $\pi K$ scattering which may also have important consequences for the neutron stars EOS. The results for the mixed pion-kaon system will be reported in future publications.

\section{acknowledgement}
The author is grateful to M. Birse and N. Walet for valuable discussions.

\end{document}